\documentstyle[amssymb,aps]{revtex}

\begin{document}
\title{Condensation and interaction range in harmonic boson traps:\ a variational
approach}
\author{J. Tempere$^1$, F. Brosens$^1$, L. F. Lemmens$^2$, J. T.\ Devreese$^{1,2}$}
\address{$^1$ Departement Natuurkunde, Universiteit Antwerpen UIA, Universiteitsplein
1, B2610 Antwerpen, Belgium.}
\address{$^2$ Departement Natuurkunde, Universiteit Antwerpen RUCA, Groenenborgerlaan
171, B2020 Antwerpen, Belgium. }
\date{15/9/1999}
\maketitle

\begin{abstract}
For a gas of $N$ bosons interacting through a two-body Morse potential a
variational bound of the free energy of a confined system is obtained. The
calculation method is based on the Feynman-Kac functional projected on the
symmetric representation. Within the harmonic approximation a variational
estimate of the effect of the interaction range on the existence of
many-particle bound states, and on the N-T phase diagram is obtained.
\end{abstract}

\draft

\section{Introduction}

In this paper, we continue the investigation proposed in \cite{TVS4} and
started in \cite{TVS5}, on the influence of realistic interactions on the
expression for the free energy of the interacting Bose system. The method we
propose is distinct from other treatments in two main aspects. Firstly, it
allows for the treatment of finite-range interatomic potentials whereas
nearly all theoretical studies relying on a mean-field description have
focused on a two-body contact potential. Secondly, the exact quantum
statistics, both of the condensate and of the non-condensate atoms, is
treated analytically at arbitrary temperature.

The experimental realization of Bose-Einstein condensation in systems of
trapped, interacting bosonic atoms \cite{AndSci269,DavPRL75,BraPRL75} has
led to renewed theoretical efforts \cite
{TVS4,TVS5,String1,Stoof1,Stoof2,String2,String3,Griffin1,Tosi1,Shi1,BayPRL96}
to understand the properties of Bose gases. The need to go beyond the
contact potential (also discussed in \cite{Burnett,Stoof3,Griffin}) appears
because the ground state energy of a contact potential with negative
scattering length (relevant for the $^7$Li) is not bounded from below. The
present study of finite-range potentials, based on a variational principle
resulting from the Jensen-Feynman inequality, avoids this artefact. This
variational principle (originally formulated by Feynman \cite{Feyn1} to
treat the problem of an electron in a polarizable medium) is extended in
section II\ to treat many-body systems with finite-range interactions,
thereby incorporating the quantum statistics of the particles analytically.
To investigate interaction potentials different from the contact potential,
the T-matrix formulation used in \cite{Stoof1,Stoof2,Stoof3} can also be
applied. In this formulation, the limit of the contact potential is given by
the long wavelength limit. In a sense, the present method is the `real
space' complement of the `momentum space' T-matrix calculation: the
knowledge of the pair correlation function $g({\bf r})$\ of the model system
in the present approach allows to effectively study the effects of the
spatial dependence of the interaction potential.

The system which we analyze in the present paper consists of a fixed number
of bosonic atoms in a parabolic confinement. The interatomic interaction
studied in detail in this paper is a Morse potential where the parameters
are determined by the scattering length and the neutral atomic radius.
Although in the most recent experiments \cite{HalTBP} mixtures of gases with
different spin states are examined, we consider only the spin-polarized Bose
gas in the present analysis. The results of the path integral variational
method elaborated in section II, applied to the spin-polarized gas of bosons
interacting through a Morse potential, are reported in section III, and the
specific case of lithium is investigated using the experimentally derived
interatomic potential (from spectroscopy measurements of Abraham {\it et al.}
\cite{AbrPRA96,AbrPRA97,AbrPRL95}). The discussion of the results and a
comparison with other methods are presented in section IV.

\section{Study of the Morse potential}

The essential property of realistic interatomic interactions is that atoms
repel at short distances and attract when they are some distance apart. In
this section, we study the effect of the Morse potential $v_2({\bf r})$,
which has these two main characteristics, on a collection of bosonic atoms: 
\begin{equation}
\frac{v_2({\bf r})}U=(1-e^{-(r-r_0)/L})^2-1,  \label{morse}
\end{equation}
where the vector ${\bf r}$ (with length $r$) is the difference in position
vectors of the two atoms, and $r_0$ is a parameter which determines the
range of the potential. $U$ is a parameter which determines the strength of
the potential, and $L$ is a parameter which determines the `stiffness' of
the potential near its minimum. These parameters are related to
experimentally observable quantities. For the Morse potential the number of
bound levels $N_{lev}$\ and the scattering length $a_{scat}$\ are given by 
\cite{Morse}: 
\begin{eqnarray}
N_{lev} &=&\left[ \frac 12\left( 1-\sqrt{8U}L\right) \right] , \\
a_{scat} &=&UL^3\frac{e^{r_0/L}-16}{4e^{-r_0/L}},
\end{eqnarray}
where the square braces\ denote the largest positive integer smaller than
the expression between brackets. The remaining parameters of (\ref{morse})
are determined by a least squares method using the shape of an
experimentally determined potential \cite{AbrPRA96,AbrPRA97,AbrPRL95}.

\subsection{The Feynman-Kac variational method}

The Feynman-Kac functional is defined as an average over a Brownian motion $%
\{R(t);t\geqslant 0\}$ with a variance that is proportional to that of
standard Brownian motion by a factor $\sqrt{\hbar /m}$, see e.g. \cite{Simon}%
. The Brownian motion provides the sample paths in a $3N$-dimensional
configuration space (whose elements $\{{\bf r}_1$,${\bf r}_2$,..,${\bf r}%
_N\} $ are denoted by $r$). The initial and final points of these paths are
incorporated in the averaging symbol $E_r$ by the index and by an indicator
function $I(R(t)-r^{\prime })$. Using these concepts, a propagator written
as 
\begin{equation}
K(r,t;r^{\prime })=E_r\left[ I(R(t)-r^{\prime })\exp \left\{ -\frac 1\hbar
\int_0^tV(R(s))ds\right\} \right]
\end{equation}
satisfies the Bloch equation for distinguishable particles 
\begin{equation}
\frac \partial {\partial t}K(r,t;r^{\prime })=\frac \hbar {2m}\nabla
^2K(r,t;r^{\prime })-\frac 1\hbar V(r)K(r,t;r^{\prime }),
\end{equation}
with 
\begin{equation}
\lim\limits_{t\downarrow 0}K(r,t;r^{\prime })=\delta (r-r^{\prime }).
\end{equation}

Using the projection techniques borrowed from \cite{FeynStatMech} and
applied to confined systems in \cite{TVS4,TVS3}, it is easy to see that the
partition function for $N$ identical particles at an inverse temperature $%
\beta =1/(k_BT)$\ is given by 
\begin{equation}
Z(N,\beta )=\frac 1{N!}\int dr\text{ }E_r\left[ \sum_P\xi ^PI(R(\beta
)-P(r))\exp \left\{ -\frac 1\hbar \int_0^\beta V(R(s))ds\right\} \right] ,
\end{equation}
where $P$ denotes permutations of the particle coordinates. A summation over
all elements of the permutation group is taken. Every permutation
contributes a factor $\xi ^P$ which is $-1$ for odd permutations of fermions
and $1$ in all other cases. If the partition function $Z_0(N,\beta )$ and
some static correlation functions of a model system, with potential energy $%
V_0$, can be calculated analytically, 
\begin{equation}
Z_0(N,\beta )=\frac 1{N!}\int dr\text{ }E_r\left[ \sum_P\xi ^PI(R(\beta
)-P(r))\exp \left\{ -\frac 1\hbar \int_0^\beta V_0(R(s))ds\right\} \right] ,
\end{equation}
this knowledge can be used to derive an upper bound for the free energy $%
F=-\log [Z(N,\beta )]/\beta ,$ relying on the Jensen-Feynman inequality: 
\begin{eqnarray}
Z(N,\beta ) &=&\frac 1{N!}\int d\bar{r}\text{ }E_{\bar{r}}\left[ \sum_P\xi
^PI\left( R(\beta )-P[r]\right) \exp \left\{ -\frac 1\hbar \int_0^\beta
V_0(R(s))ds\right\} \exp \left\{ -\frac 1\hbar \int_0^\beta
[V(R(s))-V_0(R(s))]ds\right\} \right]  \nonumber \\
&=&Z_0(N,\beta )\left\langle \exp \left[ -\frac 1\hbar \int_0^\beta \left(
V(R(s))-V_0(R(s))\right) ds\right] \right\rangle  \nonumber \\
&\geqslant &Z_0(N,\beta )\exp \left[ -\frac 1\hbar \int_0^\beta \left\langle
V(R(s))-V_0(R(s))\right\rangle ds\right] .  \label{FJ1}
\end{eqnarray}
In this expression, the angular brackets denote the quantum statistical
expectation value 
\begin{equation}
\left\langle A(R(\tau ))\right\rangle =\frac 1{Z_0(N,\beta )}\int d\bar{r}%
\text{ }E_{\bar{r}}\left[ \sum_P\xi ^PI\left( R(\beta )-P[r]\right) \exp
\left\{ -\frac 1\hbar \int_0^\beta V_0(R(s))ds\right\} A(R(\tau ))\right] .
\end{equation}

We consider a spin polarized gas of bosons interacting through a two-body
potential $v_2$ such as (\ref{morse}) and confined by an anisotropic
parabolic potential. The potential energy of this system is given by 
\begin{equation}
V=\frac m2\sum_{j=1}^N\left[ \Omega _x^2x_j^2+\Omega _y^2y_j^2+\Omega
_z^2z_j^2\right] +\sum_{j=1}^N\sum_{l=j+1}^Nv_2({\bf r}_j-{\bf r}_l),
\label{V}
\end{equation}
with $m$ the mass of the particles, and ${\bf r}_j=\{x_j,y_j,z_j\}$ the
position of the $j$-th boson. The partition function, the density and the
pair correlation function for a model system with potential energy $V_0$
given by 
\begin{equation}
V_0=\sum_{j=1}^N\frac m2\left[ \Omega _x^{\prime 2}x_j^2+\Omega _y^{\prime
2}y_j^2+\Omega _z^{\prime 2}z_j^2\right] +\sum_{j=1}^N\sum_{l=j+1}^N\frac %
\kappa 2({\bf r}_j-{\bf r}_l)^2.  \label{V0}
\end{equation}
were derived analytically in refs. \cite{TVS4,TVS3} for the isotropic case.
Substituting the real potential energy for the spin polarized gas of
interacting bosons (\ref{V}), and the potential energy of the trial system (%
\ref{V0}) in the inequality (\ref{FJ1}), one finds

\begin{eqnarray}
F &\leqslant &F_0+\sum_{j=1}^N\frac m2\left[ (\Omega _x^2-w_x^2)\left\langle
x_j^2\right\rangle +(\Omega _y^2-w_y^2)\left\langle y_j^2\right\rangle
+(\Omega _z^2-w_z^2)\left\langle z_j^2\right\rangle \right] \\
&&-\frac{N\kappa }2N\left\langle {\bf R}^2\right\rangle +\frac 12%
\left\langle \sum_{j=1}^N\sum_{l\neq j=1}^Nv_2({\bf r}_j-{\bf r}%
_l)\right\rangle ,  \nonumber
\end{eqnarray}
where $F_0=-\log [Z_0(N,\beta )]/\beta $ is the free energy of the model
system with partition function $Z_0(N,\beta )$, $w_{x,y,z}^2=(\Omega
_{x,y,z}^{\prime })^2+N\kappa /m$, and ${\bf R}=(1/N)\sum_{j=1}^N{\bf r}_j$
is the center-of-mass coordinate. Using the pair correlation function 
\begin{equation}
g({\bf r})=\frac 1{N(N-1)}\int \frac{d^3{\bf k}}{(2\pi )^3}e^{-i{\bf kr}%
}\left\langle \sum_{j=1}^N\sum_{l\neq j=1}^Ne^{i{\bf k}({\bf r}_j-{\bf r}%
_l)}\right\rangle ,
\end{equation}
the expectation value of the two-body potential\smallskip ~$v_2$ can be
rewritten 
\begin{equation}
\left\langle \sum_{j=1}^N\sum_{l=j+1}^Nv_2({\bf r}_j-{\bf r}_l)\right\rangle
=\frac{N(N-1)}2\int d{\bf r}\text{ }v_2({\bf r})g({\bf r}),
\end{equation}
and hence the variational free energy can be written as 
\begin{equation}
F\leqslant F_0+\frac m2\sum_{j=1}^N\left[ (\Omega _x^2-w_x^2)\left\langle
x_j^2\right\rangle +(\Omega _y^2-w_y^2)\left\langle y_j^2\right\rangle
+(\Omega _z^2-w_z^2)\left\langle z_j^2\right\rangle \right] -\frac{N\kappa }2%
N\left\langle {\bf R}^2\right\rangle +\frac{N(N-1)}2\int d{\bf r}\text{ }v_2(%
{\bf r})g({\bf r}).  \label{JF2}
\end{equation}
This is the central variational formula which we will use to find the
thermodynamical properties of the spin-polarized, parabolically confined,
interacting bosons. The essential role of the pair correlation function in
the evaluation of the expectation value of the interaction potential is
clear from (\ref{JF2}).

The required building blocks for the variational free energy (the
expectation values and the pair correlation function for the bosonic case),
obtained previously \cite{TVS4,TVS3} for the isotropic case, have to be
extended to the case of an anisotropic confinement potential. This
anisotropic generalization is documented in the appendix. The resulting
expression for the variational free energy is 
\begin{eqnarray}
F &\leqslant &-\frac 1\beta \log (Z_0(N,\beta ))+\sum_{i=x,y,z}\left\{ \frac %
1\beta \log \left( \frac{\sinh (\beta \hbar \Omega _i^{\prime }/2)}{\sinh
(\beta \hbar w_i/2)}\right) + 
{\displaystyle {\hbar [\Omega _i^2-(\Omega _i^{\prime })^2]\coth (\beta \hbar \Omega _i^{\prime }/2) \over 4\Omega _i^{\prime }}}%
\right\}  \nonumber \\
&&+\sum_{i=x,y,z}\frac{\hbar [\Omega _i^2-w_i^2]}{4w_i}\left[ \left(
\sum_{l=1}^N 
{\displaystyle {Z_0(N-l,\beta )\coth (\beta \hbar w_il/2) \over Z_0(N,\beta )\left[ \prod\nolimits_{j=x,y,z}2\sinh (\beta \hbar w_jl/2)\right] }}%
\right) -\coth (\beta \hbar w_i/2)\right]  \nonumber \\
&&+\frac{N(N-1)}2\int d{\bf r}\text{ }v_2({\bf r})g({\bf r}).  \label{JF3}
\end{eqnarray}

In this expression, $Z_0(N,\beta )$\ is the partition function of the model
containing $N$\ bosons at inverse temperature $\beta =1/(k_BT)$. In the
isotropic case, $\Omega _x^{\prime }=\Omega _y^{\prime }=\Omega _z^{\prime
}=\Omega ^{\prime },$ and the parameters $\Omega ^{\prime }$ and $w$ are the
variational parameters. This gives a substantial simplification, since the
variation with respect to $\Omega ^{\prime }$ can be done analytically in
the case of isotropy with $\Omega ^{\prime }=\Omega $ as the result, i.e.
the variational isotropic confinement frequency equals the isotropic
confinement frequency of the examined system. However, for the anisotropic
case, expression (\ref{JF3}) has to be minimized with respect to all four
parameters $\Omega _x^{\prime },\Omega _y^{\prime },\Omega _z^{\prime }$ and 
$\kappa $ to find the upper bound for the free energy.

\subsection{Variational free energy and condensation temperature for a Morse
potential}

The variational free energy associated with the Morse potential is found
using (\ref{JF3}). The Morse potential appears from the interparticle
interaction as the integral of the potential times the pair correlation
function: 
\begin{equation}
\frac{N(N-1)}2\int v_2({\bf r})g({\bf r})d{\bf r=}\frac{N(N-1)}2U\int \left(
e^{-2(r-r_0)/L}-2e^{-(r-r_0)/L}\right) g({\bf r})d{\bf r}  \label{vg}
\end{equation}
Denoting $P_{lj}=(1-b^j)(1-b^{l-j})/(1-b^l)$, $b=$ $\exp \{-\beta \hbar w\}$
and $a_w=\sqrt{\hbar /mw},$ and using the Fourier transform of the pair
correlation function, given in the appendix, we find for the pair
correlation function in the isotropic case: 
\begin{equation}
N(N-1)g({\bf r})=\sum_{l=2}^N\frac{Z_0(N-l,\beta )b^{3l/2}}{Z_0(N,\beta
)(1-b^l)^3}\sum_{j=1}^{l-1}\left[ 2\pi a_w^2P_{lj}(b)\right] ^{-3/2}\left[
\exp \left( -%
{\displaystyle {r^2 \over 2a_w^2}}%
{\displaystyle {1 \over P_{lj}(b)}}%
\right) +\exp \left( -%
{\displaystyle {r^2 \over 2a_w^2}}%
P_{lj}(b)\right) \right] ,  \label{g(r)}
\end{equation}
where $Z_0(N,\beta )$\ is again partition function of the model system.
Details on the anisotropic pair correlation function are given in the
appendix. In this section, we assume isotropy in order not to complicate the
formulas unnecessarily. Expression (\ref{vg}) reduces to 
\begin{eqnarray}
\frac{N(N-1)}2\int v_2({\bf r})g({\bf r})d{\bf r} &=&\frac U{2\sqrt{2\pi }}%
\sum_{l=2}^N\frac{Z_0(N-l,\beta )b^{3l/2}}{Z_0(N,\beta )(1-b^l)^3}%
\sum_{j=1}^{l-1}\left\{ e^{2r_0/L}f\left( 2P_{lj}^{1/2}\frac{a_w}L\right)
-2e^{r_0/L}f\left( P_{lj}^{1/2}\frac{a_w}L\right) \right.  \nonumber \\
&&\left. +P_{lj}^{-3}\left[ e^{2r_0/L}f\left( 2P_{lj}^{-1/2}\frac{a_w}L%
\right) -2e^{r_0/L}f\left( P_{lj}^{-1/2}\frac{a_w}L\right) \right] \right\}
\label{deltaT}
\end{eqnarray}
where $f$ is the following function of a dimensionless argument: 
\begin{equation}
f(x)=\sqrt{\frac \pi 2}\left[ 1-%
\mathop{\rm erf}%
\left( x/\sqrt{2}\right) \right] \left( x^2+1\right) e^{x^2/2}-x
\end{equation}
The expression for the variational free energy of this system is then given
by substituting (\ref{deltaT}) in (\ref{JF3}). In the isotropic case, one
finds 
\begin{eqnarray}
F_{morse} &\leqslant &-\frac{\log [Z_0(N,\beta )]}\beta +\left\{ \frac 3\beta
\log \left( \frac{\sinh (\hbar \beta \Omega /2)}{\sinh (\hbar \beta w/2)}%
\right) \right\} +\frac{3\hbar [\Omega ^2-w^2]}{4w}\left[ \left(
\sum_{l=1}^N 
{\displaystyle {Z_0(N-l,\beta )\coth (\beta \hbar wl/2) \over Z_0(N,\beta )\left[ 2\sinh (\beta \hbar wl/2)\right] ^3}}%
\right) -\coth (\beta \hbar w/2)\right]  \nonumber \\
&&+\frac{N(N-1)}2\frac U{\sqrt{2\pi }}\sum_{l=2}^N\frac{Z_0(N-l,\beta
)b^{3l/2}}{Z_0(N,\beta )(1-b^l)^3}\sum_{j=1}^{l-1}\left\{ 
\begin{array}{c}
e^{2r_0/L}f\left( 2P_{lj}^{1/2}\frac{a_w}L\right) -2e^{r_0/L}f\left(
P_{lj}^{1/2}\frac{a_w}L\right) \\ 
+P_{lj}^{-3}\left[ e^{2r_0/L}f\left( 2P_{lj}^{-1/2}\frac{a_w}L\right)
-2e^{r_0/L}f\left( P_{lj}^{-1/2}\frac{a_w}L\right) \right]
\end{array}
\right\}
\end{eqnarray}
where $\Omega $ equals the experimental confinement frequency and $w$ is the
remaining variational parameter and again $a_w=\sqrt{\hbar /mw}$. Details
for the anisotropic model can be found in the appendix. In this inequality,
the expression obtained in \cite{TVS4} for $\left\langle
\sum_{j=1}^Nr_j^2\right\rangle $ has been used.

\section{Results}

The variational free energy as function of $w$ for a system with positive
scattering length differs substantially from the case of negative scattering
length. Therefore these cases are discussed separately.

\subsection{Morse potentials with positive scattering length}

For Morse potentials with positive scattering length, $F_{morse}(w)$ has
only one minimum. This minimum is located in $0<w<\Omega $ and the minimum
value of the free energy at low temperatures is of the order of ground state
energy of the harmonic confinement potential. The variational free energy
can be used to derive the condensation temperature of the Bose gas. For this
purpose, the specific heat is calculated from the free energy ($c=-T\partial
^2F/\partial T^2$). For Morse potentials with a positive scattering length,
a peak appears in the specific heat as a function of temperature, indicating
the onset of Bose-Einstein condensation. We define the condensation
temperature as the temperature at which the specific heat reaches its
maximum. In the presence of interactions, the condensation temperature $T_c$
will differ from the condensation temperature $T_c^0$ of the non-interacting
Bose gas in the same confinement potential. This is illustrated in figure 1,
showing the specific heat as a function of temperature for several values of
the scattering length. The relative shift in the condensation temperature
induced by the interaction is denoted by $\delta _T=(T_c-T_c^0)/T_c^0$.

Figure 2 shows the interaction induced shift $\delta _T$ as a function of
the scattering length of the Morse potential. Typical parameter values for
the Morse potentials used in figure 2 are $U=33.36\times 10^9$ $\hbar \Omega 
$, $r_0=1.328\times 10^{-4}$ $a_{HO},$ $L=4.794\times 10^{-5}$ $a_{HO}$ with 
$a_{HO}=(\hbar /m\Omega )^{1/2}$. Adapting the range $r_0$ of this Morse
potential allows to change the scattering length and set it to the value one
wishes to study. Examples of scattering lengths appearing in experiments 
\cite{Scatlens,Confine} are given in table 1. A set of Morse potentials with
different scattering lengths $a_{scat}$ was constructed, and for each of
these Morse potentials, the interaction induced shift $\delta _T$ in the
condensation temperature was calculated. The results are shown in figure 2
as the full circles. It should be noted that the condensation temperature of
an ideal, trapped Bose gas and the condensation temperature of a Bose gas
interacting through a Morse potential with zero scattering length coincide.
The full line in figure 2 shows the predicted shift in condensation
temperature for a contact potential, as obtained from solving the
Gross-Pitaevski equation \cite{String2}. The results for the contact
potential and the Morse potential approach each other for small scattering
length. One could also adapt the scattering length of the Morse potential by
adapting the parameter $U$. Choosing $U$ in stead of $r_0$ as the parameter
to adjust the scattering length had no noticeable effect on the
interaction-induced shift in the condensation temperature. In the inset, the
optimal value of the variational parameter $w$ is shown as a function of
temperature. For a repulsive Morse potential ($a_{scat}/a_{HO}>0$, full
line) we find that the size of the atom cloud is expanded since $w<\Omega $.
An attractive Morse potential ($a_{scat}<0$, discussed below, dashed line in
the inset) on the other hand will contract the gas.

\subsection{Morse potentials with negative scatting length}

For Morse potentials with negative scattering length, $F_{morse}(w)$ has two
minima separated by a free energy barrier. There is again a minimum for $w$
of the order of $\Omega $, but a second minimum is found at a much higher
frequency near $w\approx \Omega (a_{HO}/r_0)^2\gg \Omega $ where $r_0$
reflects the range of the attractive part of the interaction and $a_{HO}=%
\sqrt{\hbar /m\Omega }$. This second minimum has a free energy value of the
order of $-NU$ where $U$ reflects the depth of the attractive part of the
interatomic potential. The average distance between the bosons is of the
order of $(\hbar /mw)^{-1/2}$ and is thus comparable to the range of the
interatomic potential. Hence, it is plausible that the second minimum in the
free energy, which appears for Bose gases with negative scattering length,
corresponds to a many-particle bound state. We will refer to this state as
the `clustered' state, and to the state corresponding to the minimum with $%
w\,$of the order of $\Omega $ as the `gaseous' state. Since the free energy
in the clustered state is lower than the free energy in the gaseous state,
the latter is metastable with respect to a transition to the clustered
state. When the scattering length is not negative, the minimum in the free
energy at a large (`cluster') value of $w$\ is not present. This property of
the free energy has been checked numerically for a Morse potential and has
been obtained analytically for any two-step square well potential with a
non-negative scattering length.

For the negative scattering lengths, only the specific heat associated with
the gaseous state shows a peak as a function of temperature, and thus we
find a condensation temperature only in the gaseous state, as expected. The
interaction induced shift in the condensation temperature is opposite to the
shift for potentials with positive scattering length, as shown in figure 2.

\subsubsection{Phase diagram for a Bose gas with negative scattering length}

As discussed above for the Morse potentials with negative scattering length,
there are in general two minima in the free energy. However, when the number
of particles is increased at fixed temperature, we find that the minimum in
the variational free energy associated with the gaseous state disappears
above a critical number $N_c$ of particles. The analytic expression for this
critical number at temperature zero can be found using that $Z_0(N-l,\beta
)b^{3l/2}/Z_0(N,\beta )\rightarrow 1$ and $P_{lj}\rightarrow 0$ for $%
T\rightarrow 0$, such that 
\begin{equation}
F_{morse}(T\rightarrow 0)=\frac 32\hbar w(N-1)+\frac 32\hbar \Omega +\frac{%
3\hbar (\Omega ^2-w^2)}{4w}(N-1)+\frac{N(N-1)}{\sqrt{2\pi }}Ue^{r_0/L}\left[
e^{r_0/L}f(2a_w/L)-2f(a_w/L)\right] ,
\end{equation}
Then $N_c(T\rightarrow 0)$ can be found by treating the variational equation 
$\partial F/\partial w=0$ as an equation in $w\,$and $N$ and finding the
maximal $N$ possible as a function of $w$. In the gaseous state $a_w/r_0\gg
1 $ and the asymptotic form of $f$ can be used, yielding 
\begin{equation}
N_c(T\rightarrow 0)=\frac{\sqrt{8\pi }}{5^{5/4}}\frac{a_{HO}}L\frac{\hbar
^2/(mL^2)}Ue^{-r_0/L}\left( \frac 14-\frac 4{e^{r_0/L}}\right) ,
\label{nmopo}
\end{equation}
with again $a_{HO}=\sqrt{\hbar /m\Omega }$. The relation of the Morse
potential parameters to the scattering length $a_{scat}$ of the potential
allows to rewrite (\ref{nmopo}) as $N_c(T\rightarrow 0)=-(\sqrt{8\pi }%
/5^{5/4})a_{HO}/a_{scat}\approx -0.67$ $a_{HO}/a_{scat}$. This theoretical
value for $N_c$ in the limiting case of $T\rightarrow 0$ was also obtained
by Salasnic \cite{Salas97} using a Gaussian variational wave function and
the Rayleigh-Ritz variational principle. The advantage of the present
method, however, is that unlike a variational approach based on a trial wave
function, the current technique allows to calculate $N_c$ at any
temperature. The temperature dependence of the critical number $N_c(T)$ is
shown in figure 3. Figure 3 represents a phase diagram in the $N,T$ plane
for the case of negative scattering length $a_{scat}/a_{HO}=-6.7\times
10^{-3}$, a value typical for the experiments on ultra cold trapped atoms
(see also table 1). This specific choice for $a_{scat}/a_{HO}$ corresponds
to $N_c=100$ at zero temperature. Several regions can be distinguished in
this phase diagram: a region where the metastable gaseous state exists and
is not Bose condensed$,$ a region where the metastable gaseous state exists
as Bose condensate, and finally a region where only the clustered state was
found by the present approach.

Keeping the temperature $T$ fixed, we find a number of particles $N_b(T)$
such that Bose-Einstein condensation sets in for $N_b(T)<N$. We also find
the critical number of particles $N_c(T)$ above which Bose-Einstein
condensation is no longer possible. As long as $N_b(T)<N_c(T),$
Bose-Einstein condensation can be realized at the given temperature $T$ for $%
N_b(T)<N<N_c(T)$. As the temperature is increased, $N_b(T)\rightarrow N_c(T)$
until at the tricritical temperature $T_{tc}$ (indicated in figure 2) one
has $N_b(T_{tc})=N_c(T_{tc})$. For temperatures above this tricritical
temperature, Bose-Einstein condensation is no longer possible, regardless of
the number of particles in the gas.

Keeping the number of particles $N$ fixed, we find a condensation
temperature $T_c(N)$ but also a temperature $T_{cl}(N)$ such that for
temperatures lower than $T_{cl}(N)$ only the clustered state is found by the
present approach. Thus, Bose-Einstein condensation for a given {\it total}
number of particles $N$ (both condensate and non-condensate) can only occur
in a temperature range $T_{cl}(N)<T<T_c(N)$. This means that, given a fixed
total number of particles, the gas can only be cooled to a temperature $%
T_{cl}(N)$, and not to $T=0$. Since the gas has to be cooled all the way
down to $T=0$ in order to have a condensate fraction of 100\%, and since it
can only be cooled to $T_{cl}(N),$ the maximal condensate fraction of a Bose
gas with negative scattering length will be less than 100\% for $N$ such
that $T_{cl}(N)>0$.

In our approach the maximum total number of bosons $N_c(T)$\ in the trap at
a given temperature is obtained. This number can be used to estimate the
fraction of the atoms in the condensate using $1-\{T/T_c[N_c(T)]\}^3$\ where 
$T_c(N)$\ denotes the condensation temperature for $N$\ bosons. The
condensate fraction can then be studied as a function of the temperature for
the scattering length used in the preceeding calculation, $%
a_{scat}/a_{HO}=-6.7\times 10^{-3}$. The temperature dependence of this
condensate fraction is shown in figure 4. In the inset, the number of atoms
in the condensate, calculated in ref \cite{Stoof1} within the $T$-matrix
approach but for a different scattering length $a_{scat}/a_{HO}=-4.64\times
10^{-4}$, is shown for comparison. Note that the decrease in the condensate
fraction is much less pronounced in the case of Houbiers et al. \cite{Stoof1}%
. This may be due to the fact that in the present approach the
thermodynamical properties are calculated as a function of the total number
of atoms in the trapped gas, both condensed and non-condensed. Houbiers et
al., on the other hand, calculated calculated the number of condensed atoms
in the grand canonical ensemble, given a non conserved density as a
reservoir.

As a final remark, we wish to point out that the variational method
presented in section II can also be applied to a gas of distinguishable
particles (without Bose statistics). In this manner, the effects of
statistics on for example the clustering, can be studied. For
distinguishable particles, no sums over permutations must be taken, and the
expression for the variational free energy (in isotropic confinement)
simplifies to 
\begin{eqnarray}
F_{morse}^{dist.} &\leqslant &-\frac 1\beta \log \left( 
{\displaystyle {[2\sinh (\beta \hbar \Omega /2)]^{-3} \over [2\sinh (\beta \hbar w/2)]^{3(N-1)}}}%
\right) +\frac{3\hbar (\Omega _0^2-w^2)(N-1)}4\frac{\coth (\hbar \beta w/2)}w
\nonumber \\
&&+\frac{3\hbar (\Omega _0^2-\Omega ^2)}4%
{\displaystyle {\coth (\beta \hbar \Omega /2) \over \Omega}}%
+\frac{UN(N-1)}{(2\pi )^{1/2}}\left(
-2f(a_T/L)e^{r_0/L}+f(2a_T/L)e^{2r_0/L}\right)
\end{eqnarray}
where $a_T=a_w\sqrt{\coth (\hbar \beta w/2)}$, and $\Omega _0$ is the
experimental confinement frequency. The variational free energy of the
system of distinguishable particles no longer gives rise to a peak in the
specific heat as a function of temperature, indicating that there is no
phase transition to an ordered state (the condensate). As it should be, no
condensation temperature $T_c>0$ is found. But for Morse potentials with
negative scattering length, the variational free energy will again allow for
a clustered state with respect to which the gaseous state is metastable. One
may wonder whether the critical number of particles above which the present
approach only finds the clustered state depends on the Bose statistics: is
it the same for distinguishable particles as for bosons ? The answer as
given by the present calculation, is no: the critical number of particles $%
N_c^d(T)$ for the distinguishable particles, shown as the dotted line in
figure 3, differs from the critical number of particles $N_c(T)$ for the
bosons except at $T=0$. At a given temperature, the critical number above
which only the clustered state persists is lower for bosons than for
distinguishable particles under the same conditions. For temperatures
approaching zero, the critical number for bosons approaches that for
distinguishable particles.

\subsubsection{Case study: The lithium condensate}

In contrast to a standard variational approach using a trial wave function,
there is no simple criterion to estimate how close the variational upper
bound for the free energy lies to the real value of the free energy of the
system. In order to check the robustness of the variational estimate for the
free energy, we focus on a case study for lithium. Instead of using the
Morse potential, as in the previous subsections, we use the real interaction
potential, experimentally derived from spectroscopy measurements \cite
{AbrPRA96,AbrPRA97,AbrPRL95}, and compare the results for this potential
with the analytical results found for the Morse potential.

This implies that we have to describe firstly how we used the real
interaction potential in the calculation and secondly how we obtained the
parameters of the Morse potential for these atoms. For distances less than $%
0.15$\ nm, the core-region, the real interaction potential is not determined
experimentally. We introduced a constant potential $V_{core}$\ that is
adjusted to the known scattering length \cite{AbrPRA96,AbrPRA97,AbrPRL95}.
Anticipating on the results, it should be noted that the introduction of a
constant potential does not strongly influence the results. One of the
reasons may be that the region of the core is small compared to the region
covered by a substantial value of the pair correlation. The remaining
experimental parameters of the trap, such as the confinement frequency, were
chosen in agreement with the experimental setup of \cite{BraPRL78}.

The Morse potential (\ref{morse}) for the lithium triplet interaction is
obtained as described before. The scattering length is fixed to its
experimental value ($-27.6$\ nm) and the number of bound levels (eleven, 
\cite{cotPRA94}) is introduced. The remaining parameters are determined by a
least squares fit to the measured triplet potential, leading to $%
U=33.002\times 10^9$\ $\hbar \Omega $, $r_0=1.2531\times 10^{-4}$\ $a_{HO}$, 
$L=4.5236\times 10^{-5}$\ $a_{HO}$.

The results at zero temperature are summarized in table 2. For the
variational free energy in the gaseous state, no differences were found in
the results for the Morse potential and for the experimental potential. The
variational free energy in the gaseous state, and the value of $w$
associated with it, are listed in columns 2 and 3 of table 2. In the next
two columns, the free energy and the associated $w$ are shown for the
clustered state for the Morse potential. Columns 5 and 6 show the free
energy and $w$ for the clustered state for the experimental potential. Note
that at the lowest particle numbers, the free energy of the clustered state
is increased, and even becomes larger than the free energy of the gaseous
state. This result deserves further attention, but in our opinion it is a
typical few-body problem where the molecular potentials have to be taken
into account to higher orders than we did here. Therefore we expect that the
model system is no longer adequate for such a small number of particles.

The agreement between the clustered state for the Morse potential and for
the experimental potential is good, and increases in accuracy as the number
of particles increases. Also the critical number $N_c$ remains the same. The
result for the critical number of particles for the experimental setup of 
\cite{BraPRL78}, with scattering length and trapping parameters such that $%
a_{scat}/a_{HO}=4.64\times 10^{-4}$, is $N_c=1443$ in the present approach.
This number $N_c$\ represents the maximum number of atoms in the trap that
still can undergo Bose-Einstein condensation. Thus it is an upper bound to
the average number of atoms which is reported in \cite{BraPRL78} to lie
between $600-1300$.

\section{Discussion and Conclusion}

The method proposed in this paper allows to incorporate any two-body
interaction $v_2$ in the description of the thermodynamics of an interacting
Bose gas. From this point of view, its application is more general than the
Gross-Pitaevskii model which takes into account a two-body contact potential 
$v_\delta ({\bf r})=4\pi \delta ({\bf r})\hbar ^2a_{scat}/m$ with scattering
length $a_{scat}$, and which hence neglects the range and shape of the
interatomic potential. From the expression of the expectation value of the
two-body interaction in the Jensen-Feynman inequality, it is clear that this
approximation will be valid as long as the range of the interatomic
potential (of the order of $r_0$\ for the Morse potential used above) is
much shorter than the typical length scale $\xi $ over which the pair
correlation function varies. Indeed, in that case, the expectation value of
the interatomic potential essentially depends on the value of the pair
correlation function in the origin and on the scattering length: 
\begin{equation}
\left\langle v_2\right\rangle =\frac{N(N-1)}2\int d{\bf r}\text{ }g({\bf r}%
)v_2({\bf r})\longrightarrow \left\langle v_2\right\rangle _{r_0\ll \xi
}\approx \frac{N(N-1)}2g(0)\int d{\bf r}\text{ }v_2({\bf r})=\frac{2\pi
\hbar ^2a_{scat}}mN(N-1)g(0).
\end{equation}
However, even for short-range interactions, the typical length scale of the
pair correlation function can become comparable to the range of the
interatomic interactions. This happens for interatomic potentials which
allow for a clustered state (a bound state involving many particles). In
that case, the details of the interaction potential become important and
models based on the Gross-Pitaevskii equation using a contact potential lead
to an unphysical result: the free energy in these models is not bound from
below. This artefact is avoided in the present treatment by taking into
account the range of the interatomic potential.

Bose-Einstein condensation in a gas with negative scattering length has been
achieved experimentally: for a parabolically confined gas of lithium atoms a
Bose-Einstein condensate of at most $600-1300$ bosons \cite{BraPRL78} was
observed, comparable to the critical number for this experiment, predicted
from the theory presented here. It should be noted that the present analysis
does not describe the dynamics of a cooling Bose gas nor does it describe
the dynamics of the formation of the clustered state. The critical number of
bosons beyond which Bose-Einstein condensation is no longer possible may be
different in the experiment due to the dynamical effects.

In the theoretical approach presented here, as well as in the two-fluid
models \cite{String2,String3,Griffin1,Tosi1,Shi1} for the condensate and
even in the Monte Carlo simulations \cite{KraPRL96,Gruunp97}, it is tacitly
assumed that the system is in thermal equilibrium. It is also clear that the
measurements are done on systems that are not in equilibrium at all; even
obtaining a steady state seems at present out of the experimental reach \cite
{WilPRL98,HulAPB65}. This means that in discussing the theoretical results
we rely on our intuition of how the calculated equilibrium situation is
reached without conflicting with the known experimental facts for the
non-equilibrium situation. For our approach, this means that we have to
argue how it is possible that a metastable state with a much higher energy
than the ground state can be studied experimentally, while the ground state
itself does not even show up circumstantially. For distinguishable particles
it is generally accepted that the life-time of a metastable state is
proportional to the energy-barrier between the two states and inversely
proportional to their energy difference. There are different scenarios
possible. The first one implicates gravity: as soon as a cluster of a few
atoms is formed it leaves the trap due to the gravitational force, thereby
reducing the number of atoms available for condensation and increasing the
average energy present in the trap, because the low-energy states can escape
now. In this scenario one is allowed to start with a number of atoms larger
than the calculated $N_c$. After some time of operation, clustering will
reduce the number of atoms in the trap. Once the number of atoms is lower,
the metastable state becomes available as the first excited state of the
system. If the collision rate in this state does not differ fundamentally
from the collision rate of the previous situation, the clustering will go
on, and eventually all particles will have left the trap. If there is a
change in collision rate, or if the state is almost collision free, the
metastable state can get the necessary life-time to be available for
experimentation. For $^{87}$Rb this change in collision-rate was found \cite
{BuPRL97} to be due to the influence of the condensate on the loss-rate of
three-body recombinations. This characteristic was predicted by Kagan {\it %
et al.} \cite{KagJETP85} who suggests also to use it as the signature of the
presence of the condensate. The physical origin of the change in collision
rate is due to a change in distribution. While in the high temperature phase
the distribution is Gaussian (at least to a good approximation) it would
remain Gaussian (with different parameters) if the particles were
distinguishable. Projecting it on the symmetric irreducible representation
of the permutation group as required for indistinguishable particles, the
distribution takes the equilibrium of the condensate and its excitations
into account.

In this paper, we presented a path-integral variational method which 1)
exactly treats the quantum statistics at any temperature and 2) allows to
incorporate finite range two-body interactions in the description of the
thermodynamics of the Bose gas. After presenting this technique, we focused
on a gas of parabolically trapped bosonic atoms interacting though a Morse
potential. The effect of this interaction on the condensation temperature
was studied. For Morse potentials with a negative scattering length, a phase
diagram was derived showing the influence of the interaction on the region
in $N,T$ parameter space favorable to Bose-Einstein condensation. Apart from
the interaction induced shift in condensation temperature, the region in $%
N,T $ parameter space available for Bose-Einstein condensation is reduced by
the presence of a clustered phase. The effect of the interaction and the
Bose statistics on the critical number of particles above which the gaseous
state was no longer found in the present treatment, was calculated. We
identified a temperature above which\ Bose-Einstein condensation is no
longer possible regardless of the number of particles.

\section*{Acknowledgments}

Discussions with S. Stringari, Y. Kagan, S. Giorgini and L. Salasnic are
gratefully acknowledged. We thank H. Stoof for discussions and for making
the results on the $^7$Li--potential available. This work is performed
within the framework of the FWO\ projects No. 1.5.729.94, 1.5.545.98,
G.0287.95, G.0071.98 and WO.073.94N (Wetenschappelijke
Onderzoeksgemeenschap, Scientific Research Community of the FWO on ``Low
Dimensional Systems''), the ``Interuniversitaire Attractiepolen -- Belgische
Staat, Diensten van de Eerste Minister -- Wetenschappelijke, Technische en
Culturele Aangelegenheden'', and in the framework of the BOF\ NOI 1997
projects of the Universiteit Antwerpen. Two of the authors (J.T., aspirant
bij het Fonds voor Wetenschappelijk Onderzoek - Vlaanderen, and F.B.)
acknowledge the FWO (Fonds voor Wetenschappelijk Onderzoek-Vlaanderen) for
financial support.

\section*{Appendix}

The expectation values of one-body and two-body potential energies appearing
in the Jensen-Feynman inequality can be expressed as a function of the
density 
\begin{equation}
\sum_{j=1}^Nv_1({\bf r}_j)=\int d{\bf r}\text{ }v_1({\bf r})n({\bf r})\text{
with }n({\bf r})=\int \frac{d{\bf k}}{(2\pi )^3}e^{-i{\bf kr}}\left\langle
\sum_{j=1}^Ne^{i{\bf kr}_j}\right\rangle ,
\end{equation}
and of the pair correlation function: 
\begin{eqnarray}
\frac 12\sum_{j=1}^N\sum_{l=1}^Nv_2({\bf r}_j-{\bf r}_l)\Theta (l &\neq &j)=%
\frac{N(N-1)}2\int d{\bf r}\text{ }v_2({\bf r})g({\bf r})\, \\
\text{with }g({\bf r}) &=&\frac 1{N(N-1)}\int \frac{d{\bf k}}{(2\pi )^3}e^{-i%
{\bf kr}}\left\langle \sum_{j=1}^N\sum_{l=1}^Ne^{i{\bf k}\left( {\bf r}_j-%
{\bf r}_l\right) }\Theta (l\neq j)\right\rangle .
\end{eqnarray}
The symbol $\Theta (expr)=1$ if the logical expression $expr$ is true$,$ and
zero otherwise. These quantities have been discussed in detail in \cite{TVS4}%
. In this appendix, the generalization of the results of \cite{TVS4} to the
case of anisotropic confinement will be given. For a spin-polarized gas of
bosons in an anisotropic parabolic confinement potential with frequencies $%
\Omega _x,\Omega _y,\Omega _z$ and a two-body potential energy $(\kappa
/4)\sum_{j,l}({\bf r}_j-{\bf r}_l)^2,$ it is given by 
\begin{eqnarray}
K({\bf r}_1,..,{\bf r}_N;\beta |{\bf r}_1,..,{\bf r}_N;0) &=&\frac 1{N!}C(%
{\bf R})\sum_P%
\mathop{\displaystyle \prod }%
\limits_{j=1}^NK_{w_x,w_y,w_z}(P[{\bf r}_j];\beta |{\bf r}_j;0), \\
C({\bf R}) &=&\frac{K_{\Omega _x,\Omega _y,\Omega _z}(\sqrt{N}{\bf R};\beta |%
\sqrt{N}{\bf R};0)}{K_{w_x,w_y,w_z}(\sqrt{N}{\bf R};\beta |\sqrt{N}{\bf R};0)%
},
\end{eqnarray}
where the first sum runs over all particle permutations $P$; $w_i=(\Omega
_i^2-N\kappa )^{1/2}$ are the renormalized frequencies; ${\bf R}=(1/N)\sum_j%
{\bf r}_j$ is the center of mass coordinate; and $K_{v_x,v_y,v_z}$ is the
single-particle path integral propagator for a anisotropic harmonic
oscillator with frequencies $v_x,v_y,v_z$. Hence 
\begin{eqnarray}
Z_0(N,\beta ) &=&\sum_{j=1}^N\int d{\bf r}_1..d{\bf r}_N\text{ }C({\bf R})%
\frac 1{N!}\sum_P\left[ 
\mathop{\displaystyle \prod }%
\limits_{v=1}^NK_{w_x,w_y,w_z}(P[{\bf r}_v];\beta |{\bf r}_v;0)\right] , \\
\left\langle \sum_{j=1}^Ne^{i{\bf kr}_j}\right\rangle &=&\frac 1{Z_0(N,\beta
)}\sum_{j=1}^N\int d{\bf r}_1..d{\bf r}_N\text{ }C({\bf R})\frac 1{N!}%
\sum_P\left[ 
\mathop{\displaystyle \prod }%
\limits_{v=1}^NK_{w_x,w_y,w_z}(P[{\bf r}_v];\beta |{\bf r}_v;0)e^{i{\bf kr}%
_j}\right] , \\
\left\langle \sum_{j=1}^N\sum_{l=1}^Ne^{i{\bf k}\left( {\bf r}_j-{\bf r}%
_l\right) }\Theta (l\neq j)\right\rangle &=&\frac 1{Z_0(N,\beta )}%
\sum_{j=1}^N\sum_{l=1}^N\Theta (l\neq j)\int d{\bf r}_1..d{\bf r}_N\text{ }C(%
{\bf R})\frac 1{N!}  \nonumber \\
&&\times \sum_P\left[ 
\mathop{\displaystyle \prod }%
\limits_{v=1}^NK_{w_x,w_y,w_z}(P[{\bf r}_v];\beta |{\bf r}_v;0)e^{i{\bf k}%
\left( {\bf r}_j-{\bf r}_l\right) }\right] ,
\end{eqnarray}
where $Z_0(N,\beta )$ indicates the partition function in the analytical
model system. Each permutation can be written as a cyclic decomposition with 
$M_1$ cycles of length $1$, $M_2$ cycles of length $2$,... The sum over
permutations can be transformed into a sum over cyclic decompositions, as in
Feynman's treatment of the ideal homogeneous Bose gas \cite{FeynStatMech}: 
\begin{equation}
\sum_P\rightarrow \sum_{\{M_1,M_2,..\}}N!%
\mathop{\displaystyle \prod }%
\limits_l\frac 1{M_l!l^{M_l}}\Theta \left( \sum_llM_l=N\right) .
\end{equation}
The constraint in the cyclic sum appears in order to guarantee that the
total number of elements in all cycles of the cyclic decomposition equals
the total number of bosons in the gas. For example, the partition sum
becomes 
\begin{eqnarray}
Z_0(N,\beta ) &=&\sum_{j=1}^N\int d{\bf r}_1..d{\bf r}_N\text{ }C({\bf R}%
)\sum_{\{M_1,M_2,..\}}\left[ 
\mathop{\displaystyle \prod }%
\limits_l\frac 1{M_l!l^{M_l}}K_l^{M_l}\right] \Theta \left(
\sum_llM_l=N\right) , \\
K_l &=&K_{w_x,w_y,w_z}({\bf r}_1;\beta |{\bf r}_l;0)...K_{w_x,w_y,w_z}({\bf r%
}_3;\beta |{\bf r}_2;0)K_{w_x,w_y,w_z}({\bf r}_2;\beta |{\bf r}_1;0).
\end{eqnarray}
The restriction $\sum_llM_l=N$ is prohibitive for performing the summation
directly. To lift this restriction, and perform the summation, the
generating functions are introduced: 
\begin{eqnarray}
\Xi (u) &=&\sum_{N=0}^\infty Z_0(N,\beta )u^N,  \label{xi} \\
G_1({\bf k},u) &=&\sum_{N=0}^\infty Z_0(N,\beta )\left\langle
\sum_{j=1}^Ne^{i{\bf kr}_j}\right\rangle u^N,  \label{g1} \\
G_2({\bf k},u) &=&\sum_{N=0}^\infty Z_0(N,\beta )\left\langle
\sum_{j=1}^N\sum_{l=1}^Ne^{i{\bf k}\left( {\bf r}_j-{\bf r}_l\right) }\Theta
(l\neq j)\right\rangle u^N.  \label{g2}
\end{eqnarray}
For these generating functions, the summations and integrations can be
performed analytically as for the isotropic case \cite{TVS4,TVS3}. Using the
notations $b_i=\exp \{-\beta \hbar w_i\}$ for $i=x,y,z$ and $\beta =1/k_BT$,
the generating functions are given by 
\begin{eqnarray}
\Xi (u) &=&\exp \left\{ \sum_{l=1}^N\frac{u^l(b_xb_yb_z)^{l/2}}{%
(1-b_x^l)(1-b_y^l)(1-b_z^l)}\right\} , \\
G_1({\bf k},u) &=&\Xi (u)\sum_{l=1}^N\frac{u^l(b_xb_yb_z)^{l/2}}{%
(1-b_x^l)(1-b_y^l)(1-b_z^l)}\exp \left\{ -\sum_{i=x,y,z}\frac{\hbar k_i^2}{%
4mw_i}\frac{1+b_i^l}{1-b_i^l}\right\} , \\
G_2({\bf k},u) &=&\Xi (u)\sum_{l=2}^N\frac{u^l(b_xb_yb_z)^{l/2}}{%
(1-b_x^l)(1-b_y^l)(1-b_z^l)}  \nonumber \\
&&\times \sum_{j=1}^{l-1}\left[ \exp \left( -\sum_{i=x,y,z}\frac{\hbar k_i^2%
}{2mw_i}P_{l,j}(b_i)\right) +%
\mathop{\displaystyle \prod }%
\limits_{i=x,y,z}P_{l,j}^{-1}(b_i)\exp \left( -\frac{\hbar k_i^2}{2mw_i}%
\frac 1{P_{l,j}(b_i)}\right) \right] ,
\end{eqnarray}
with $P_{l,j}(b)=(1-b^j)(1-b^{l-j})/(1-b^l)$. Two schemes have been used to
retrieve the expectation values from their generating function. The
contour-integration scheme \cite{Kubo} was applied in \cite{TVS6}, and
yields 
\begin{eqnarray}
Z_0(N,\beta ) &=&\frac 1{N!}\left. \frac{d^N\Xi (u)}{du^N}\right| _{u=0}=%
\frac 1{2\pi }%
\displaystyle \int %
\limits_0^{2\pi }\frac{\Xi (ue^{i\theta })}{u^N}e^{-iN\theta }d\theta ,
\label{aa} \\
\left\langle \sum_{j=1}^Ne^{i{\bf kr}_j}\right\rangle &=&\frac 1{Z_0(N,\beta
)}\frac 1{N!}\left. \frac{d^NG_1(u)}{du^N}\right| _{u=0}=\frac{\Xi (u)/u^N}{%
Z_0(N,\beta )}\frac 1{2\pi }%
\displaystyle \int %
\limits_0^{2\pi }\frac{G_1(ue^{i\theta })}{\Xi (u)}e^{-iN\theta }d\theta , \\
\left\langle \sum_{j=1}^N\sum_{l=1}^Ne^{i{\bf k}\left( {\bf r}_j-{\bf r}%
_l\right) }\Theta (l\neq j)\right\rangle &=&\frac 1{Z_0(N,\beta )}\frac 1{N!}%
\left. \frac{d^NG_2(u)}{du^N}\right| _{u=0}=\frac{\Xi (u)/u^N}{Z_0(N,\beta )}%
\frac 1{2\pi }%
\displaystyle \int %
\limits_0^{2\pi }\frac{G_2(ue^{i\theta })}{\Xi (u)}e^{-iN\theta }d\theta .
\label{cc}
\end{eqnarray}
In principle, any value can be chosen for $u$. Fastest numerical convergence
however is achieved for the steepest descent solution, when $u$ is chosen
such that $N=-\beta ^{-2}\partial (\ln \Xi (u))/\partial (\log u)$. It would
be incorrect to put the expectation value equal to the generating function
itself, in which the steepest descent value $u=e^{\beta \mu }$ for the
parameter $u$ in the free energy generating function, is substituted.

The second method is to collect the coefficient of $u^N$. This results in
recursion relations between the expectation value for $N$ particles and the
expectation values for smaller number of particles: 
\begin{eqnarray}
Z_0(N,\beta ) &=&\sum_{l=1}^N\frac{(b_xb_yb_z)^{l/2}}{%
(1-b_x^l)(1-b_y^l)(1-b_z^l)}Z_0(N-l,\beta )\text{ with }Z_0(0,\beta )=1
\label{bb} \\
\left\langle \sum_{j=1}^Ne^{i{\bf kr}_j}\right\rangle &=&\sum_{l=1}^N\frac{%
Z_0(N-l,\beta )(b_xb_yb_z)^{l/2}}{Z_0(N,\beta )(1-b_x^l)(1-b_y^l)(1-b_z^l)}%
\exp \left\{ -\sum_{i=x,y,z}\frac{\hbar k_i^2}{4mw_i}\frac{1+b_i^l}{1-b_i^l}%
\right\} \\
\left\langle \sum_{j=1}^N\sum_{l=1}^Ne^{i{\bf k}\left( {\bf r}_j-{\bf r}%
_l\right) }\Theta (l\neq j)\right\rangle &=&\sum_{l=2}^N\frac{Z_0(N-l,\beta
)(b_xb_yb_z)^{l/2}}{Z_0(N,\beta )(1-b_x^l)(1-b_y^l)(1-b_z^l)}  \nonumber \\
&&\times \sum_{j=1}^{l-1}\left[ \exp \left( -\sum_{i=x,y,z}\frac{\hbar k_i^2%
}{2mw_i}P_{l,j}(b_i)\right) +%
\mathop{\displaystyle \prod }%
\limits_{i=x,y,z}P_{l,j}^{-1}(b_i)\exp \left( -\frac{\hbar k_i^2}{2mw_i}%
\frac 1{P_{l,j}(b_i)}\right) \right]  \label{dd}
\end{eqnarray}
The numerical implementation of the recursion relations gives high accuracy
at the cost of computing time. In practice, they become too time-consuming
for more than a few thousand particles, and expressions (\ref{aa})-(\ref{cc}%
) have then to be computed. In summary, the expectation values in the
path-integral formalism were calculated by introducing the generating
functions (\ref{xi})-(\ref{g2}) for these expectation values. By
constructing the generating functions, one can analytically perform the sum
over all permutations, necessary to incorporate the quantum statistics. The
expectation value can then be extracted from the generating function using
the inversion formulas (\ref{aa})-(\ref{cc}) or (\ref{bb})-(\ref{dd}).

\bigskip \newpage

\section*{Figure Captions}

{\bf Figure 1:}

The specific heat of a parabolically trapped Bose gas of 2000 atoms is shown
as a function of temperature for several scattering lengths. In the inset
the relative shift in the condensation temperature induced by the
interparticle interaction is shown as a function of the scattering length.
The condensation temperature is determined from the maximum of the specific
heat. The full line is derived using Gross-Pitaevskii theory \cite{String2},
the filled circles are obtained using the approach presented here.

\medskip

{\bf Figure 2:}

The interaction induced shift of the condensation temperature is shown as a
function of the interaction strength $a_{scat}/a_{HO}$ for Morse potentials
(filled circles, calculated in the present approach) and contact potentials
(full line, calculated from the Gross-Pitaevskii equation \cite{String2}).
In the inset, the variational optimal value for the parameter $w/\Omega $ is
shown as a function of temperature, for a repulsive (full line) and an
attractive (dashed line) Morse potential and for the non-interacting,
parabolically trapped Bose gas (dotted line).

\medskip

{\bf Figure 3:}

A $(N,T)$--phase diagram is shown for a gas of bosons interacting through a
Morse potential with negative scattering length $a_{scat}/a_{HO}=-0.0067$
with $a_{HO}=\sqrt{\hbar /m\Omega }$, such that $N_c(T=0)=100$. The dashed
line shows the condensation temperature $T_c$ as a function of the number of
bosons. The full line shows the critical number of bosons $N_c$ beyond which
the gaseous state no longer exists, as a function of temperature, and the
dotted line shows this critical number for distinguishable particles under
the same conditions. The temperature $T_{tc}$\ above which BEC\ is not
possible regardless of the number of bosons, is indicated: $T_{tc}=9.43$\ $%
\hbar \Omega /k_B$, $N_c(T_{tc})=1363$.

\medskip

{\bf Figure 4:}

$N_c$\ is the maximum number of atoms in the trap which can undergo
Bose-Einstein condensation. Using this number, and the associated
condensation temperature $T_c(N_c)$\ the maximal fraction of atoms in the
condensate is shown as a function of temperature for a gas with scattering
length $a_{scat}/a_{HO}=-6.7\times 10^{-3}$. In the inset, the maximum
occupation of the condensate calculated in \cite{Stoof1} with a T-matrix
approach and for a scattering length $a_{scat}/a_{HO}=-4.6\times 10^{-4}$\
is shown.

\bigskip \newpage

\section*{Tables}

{\bf Table 1}:

Typical values for system parameters in experiments on ultra cold Bose
gases. For a set of chosen alkali atoms we list the scattering lengths $%
a_{scat}$ (taken from \cite{Scatlens}), the frequencies of the parabolic
confinement potential (taken from \cite{Confine}), typical number of atoms
for which Bose-Einstein condensation is observed (where applicable), the
effective interaction strength given by $a_{scat}/a_{HO}$ and the critical
number $N_c$ (found by the present method) at zero temperature (only
applicable for negative scattering lengths). `n.a.' stands for `not
applicable'.

\begin{tabular}{|llllcl|}
\hline
Atom & $a_{scat}$ & $N$ & $\tilde{\Omega}=\sqrt[3]{\Omega _x\Omega _y\Omega
_z}$ & $a_{scat}/a_{HO}$ & $N_c$ \\ \hline
$^7$Li & $-1.44\pm 0.04$ nm & 1300 & \multicolumn{1}{c}{144 Hz} & $%
\allowbreak -4.64\times 10^{-4}$ & $1443$ \\ 
$^{23}$Na, $\left| 1,-1\right\rangle $ & $4.9\pm 1.4$ nm & 500000 & 
\multicolumn{1}{c}{416 Hz} & $\allowbreak 0.0048$ & n.a. \\ 
$^{87}$Rb & $4.6\pm 1.1$ nm & 4500 & \multicolumn{1}{c}{187 Hz} & $0.0058$ & 
n.a. \\ 
$^{85}$Rb & $-50$ nm $<$ a$_{scat}<-3$ nm & n.a. & \multicolumn{1}{c}{%
\symbol{126}1 Hz} & $-0.0050$ & $134$ \\ 
$^{133}$Cs & $-13$ nm & n.a. & \multicolumn{1}{c}{34 Hz} & $-0.008\,7$ & $77$
\\ \hline
\end{tabular}

\medskip

{\bf Table 2:}

Results of the variation for the free energy at temperature zero, in units $%
m_{\text{Li}}=\hbar =\Omega =1$. The first column shows the number of
particles in the parabolic confinement. The next columns show the optimal
variational value for the parameter $w$ and the variational free energy for
(i) the gaseous state (identical results were found for the Morse potential
and the real potential), (ii) the Morse potential and (iii) the
experimentally derived triplet potential \cite{AbrPRA96,AbrPRA97,AbrPRL95}.
Comments are added in the last column.

\begin{tabular}{|l|rc|rl|rl|c|}
\hline
& Gaseous &  & Clustered & (Morse) & Clustered & (Exact) & comment \\ 
$N$ & \multicolumn{1}{|c}{$w_1$} & $F_1/N$ & \multicolumn{1}{|c}{$%
w_{2,morse} $} & \multicolumn{1}{c|}{$F_{2,morse}/N$} & \multicolumn{1}{|c}{$%
w_{2,lith}$} & \multicolumn{1}{c|}{$F_{2,lith}/N$} &  \\ \hline
\multicolumn{1}{|l|}{2} & \multicolumn{1}{|c}{1.00037} & \multicolumn{1}{l|}{
1.49981} & \multicolumn{1}{|c}{no minimum} & \multicolumn{1}{c|}{} & 
\multicolumn{1}{|c}{no minimum} & \multicolumn{1}{c|}{} & only \\ 
\multicolumn{1}{|l|}{5} & \multicolumn{1}{|c}{1.00091} & \multicolumn{1}{l|}{
1.49926} & \multicolumn{1}{|c}{no minimum} & \multicolumn{1}{c|}{} & 
\multicolumn{1}{|c}{no minimum} & \multicolumn{1}{c|}{} & gaseous \\ 
\multicolumn{1}{|l|}{10} & \multicolumn{1}{|c}{1.00184} & 
\multicolumn{1}{l|}{1.49833} & \multicolumn{1}{|c}{no minimum} & 
\multicolumn{1}{c|}{} & \multicolumn{1}{|c}{no minimum} & 
\multicolumn{1}{c|}{} & state \\ 
\multicolumn{1}{|l|}{20} & \multicolumn{1}{|c}{1.00373} & 
\multicolumn{1}{c|}{1.49647} & \multicolumn{1}{|c}{no minimum} & 
\multicolumn{1}{c|}{} & \multicolumn{1}{|c}{no minimum} & 
\multicolumn{1}{c|}{} &  \\ \cline{1-1}\hline
\multicolumn{1}{|l|}{21} & \multicolumn{1}{|c}{1.00392} & 
\multicolumn{1}{c|}{1.49628} & \multicolumn{1}{|c}{53389.8} & 
\multicolumn{1}{c|}{4714.57} & \multicolumn{1}{|c}{57555.7} & 
\multicolumn{1}{c|}{4268.57} & clustered \\ 
\multicolumn{1}{|l|}{22} & \multicolumn{1}{|c}{1.00411} & 
\multicolumn{1}{c|}{1.49610} & \multicolumn{1}{|c}{60304.0} & 
\multicolumn{1}{c|}{3005.24} & \multicolumn{1}{|c}{63920.3} & 
\multicolumn{1}{c|}{2406.39} & state is \\ 
\multicolumn{1}{|l|}{23} & \multicolumn{1}{|c}{1.00430} & 
\multicolumn{1}{c|}{1.49591} & \multicolumn{1}{|c}{65178.6} & 
\multicolumn{1}{c|}{1100.04} & \multicolumn{1}{|c}{68640.5} & 
\multicolumn{1}{c|}{357.693} & metastable \\ \hline
\multicolumn{1}{|l|}{24} & \multicolumn{1}{|c}{1.00449} & 
\multicolumn{1}{c|}{1.49572} & \multicolumn{1}{|c}{69042.2} & 
\multicolumn{1}{c|}{--948.329} & \multicolumn{1}{|c}{72454.2} & 
\multicolumn{1}{c|}{--1831.68} & gaseous \\ 
\multicolumn{1}{|l|}{100} & \multicolumn{1}{|c}{1.01925} & 
\multicolumn{1}{l|}{1.48140} & \multicolumn{1}{|c}{111411.} & 
\multicolumn{1}{l|}{--2.21442 10$^5$} & \multicolumn{1}{|c}{115733.} & 
\multicolumn{1}{c|}{--2.34133 10$^5$} & state is \\ 
\multicolumn{1}{|l|}{1000} & \multicolumn{1}{|c}{1.31950} & 
\multicolumn{1}{l|}{1.27743} & \multicolumn{1}{|c}{120331.} & 
\multicolumn{1}{l|}{--3.01629 10$^6$} & \multicolumn{1}{|c}{124993.} & 
\multicolumn{1}{c|}{--3.17466 10$^6$} & metastable \\ 
\multicolumn{1}{|l|}{1443} & \multicolumn{1}{|c}{2.16973} & 
\multicolumn{1}{l|}{1.11880} & \multicolumn{1}{|c}{120623.} & 
\multicolumn{1}{l|}{--4.39386 10$^6$} & \multicolumn{1}{|c}{125297.} & 
\multicolumn{1}{c|}{--4.62399 10$^6$} &  \\ \hline
\multicolumn{1}{|l|}{1444} & \multicolumn{1}{|c}{no minimum} & 
\multicolumn{1}{l|}{} & \multicolumn{1}{|c}{120624.} & \multicolumn{1}{l|}{
--4.39697 10$^6$} & \multicolumn{1}{|c}{125298.} & \multicolumn{1}{c|}{
--4.62726 10$^6$} & only clustered \\ 
\multicolumn{1}{|l|}{1500} & \multicolumn{1}{|c}{no minimum} & 
\multicolumn{1}{l|}{} & \multicolumn{1}{|c}{120648.} & \multicolumn{1}{l|}{
--4.57111 10$^6$} & \multicolumn{1}{|c}{125323.} & \multicolumn{1}{c|}{
--4.81048 10$^6$} & state \\ \hline
\end{tabular}

\medskip

{\bf Table 3}:

In this table, a description of the different length scales used in the
paper is given.

\begin{tabular}{|l|l|}
\hline
symbol & description \\ \hline
\multicolumn{1}{|c|}{$a_{scat}$} & scattering length \\ 
\multicolumn{1}{|c|}{$a_{HO}$} & $\sqrt{\hbar /(m\Omega )}$ \\ 
\multicolumn{1}{|c|}{$a_w$} & $\sqrt{\hbar /(mw)}$ \\ 
\multicolumn{1}{|c|}{$a_T$} & $a_w\sqrt{\coth [\hbar w/(2k_BT)]}$ \\ 
\multicolumn{1}{|c|}{$r_0$} & distance at which the Morse potential \\ 
& reaches its minimum \\ \hline
\end{tabular}


\begin{references}
\bibitem{TVS4}  F. Brosens, J. T.\ Devreese, and L. F. Lemmens, Phys. Rev. E 
{\bf 55}, 6795 (1997).

\bibitem{TVS5}  J. Tempere, F. Brosens, L. F. Lemmens, and J. T. Devreese,
Solid\ State Commun. {\bf 107}, 51 (1998).

\bibitem{AndSci269}  M. H. Anderson, J. R. Ensher, M. R. Matthews, C. E.
Wieman, and E. A. Cornell, Science {\bf 269}, 198 (1995).

\bibitem{DavPRL75}  K. B. Davis, M.-O. Mewes, M. R. Andrews, N. J. Van
Druten, D. S. Durfee, D. M. Kurn, W. Ketterle, Phys. Rev. Lett. {\bf 75},
3969 (1995).

\bibitem{BraPRL75}  C. C. Bradley, C. A. Sackett, J. J. Tollett, and R. G.
Hulet, Phys. Rev. Lett. {\bf 75}, 1687 (1995).

\bibitem{String1}  F. Dalfovo and S. Stringari, Phys. Rev. A {\bf 53}, 2477
(1996).

\bibitem{Stoof1}  M. Houbiers and H. T. C. Stoof, Phys. Rev. A {\bf 54},
5055 (1996).

\bibitem{Stoof2}  C. A. Sackett, H. T. C. Stoof, and R. G. Hulet, Phys. Rev.
Lett. {\bf 80}, 2031 (1998).

\bibitem{String2}  S. Giorgini, L. P. Pitaevskii, and S. Stringari, Phys.
Rev. A {\bf 54}, 4633 (1996).

\bibitem{String3}  S. Giorgini, L. P. Pitaevskii, and S. Stringari, Phys.
Rev. Lett. {\bf 78}, 3987 (1997).

\bibitem{Griffin1}  D. A. W. Hutchinson, E. Zaremba, and A. Griffin, Phys.
Rev. Lett. {\bf 78}, 1842 (1997).

\bibitem{Tosi1}  A. Minguzzi, S. Conti, and M. P. Tosi, J. Phys. Cond. Mat. 
{\bf 9}, L33 (1997).

\bibitem{Shi1}  H. Shi and W.-M. Zheng, Phys. Rev. A {\bf 56}, 1046 (1997).

\bibitem{BayPRL96}  G. Baym and C. Pethick, Phys. Rev. Lett. {\bf 76}, 6
(1996).

\bibitem{Burnett}  N. P. Proukakis and K. Burnett, J. Res. Natl. Inst.
Stand. Technol. {\bf 101}, 457 (1996).

\bibitem{Stoof3}  M. Bijlsma and H. T. C. Stoof, Phys.\ Rev. A {\bf 55}, 498
(1997).

\bibitem{Griffin}  A. Griffin, Phys. Rev. B {\bf 53}, 9341 (1996).

\bibitem{Feyn1}  R. P. Feynman, Phys. Rev. {\bf 97}, 660 (1955).

\bibitem{HalTBP}  D. S. Hall, M. R. Matthews, C. E. Wieman and E. A.
Cornell, Phys. Rev. Lett. {\bf 81}, 1543 (1998).

\bibitem{AbrPRA96}  E. R. I.\ Abraham, W. I. McAlexander, H. T. C. Stoof,
and R. G. Hulet, Phys. Rev. A {\bf 53}, 3092 (1996).

\bibitem{AbrPRA97}  E. R. I.\ Abraham, W. I. McAlexander, J. M. Gerton, R.
G. Hulet, R. Cot\'{e}, and A. Dalgarno, Phys. Rev. A {\bf 55,} 3299 (1997).

\bibitem{AbrPRL95}  E. R. I. Abraham, W. I. McAlexander, C. A. Sackett, and
R. G. Hulet, Phys. Rev. Lett. {\bf 74}, 1315 (1995).

\bibitem{Morse}  I. M. Torrens, Interatomic Potentials (Academic Press, New
York, 1972); L. P. Pitaevskii, in Proceedings of the 1998 ``Enrico Fermi''
Int. School, Varenna\ (to be published).

\bibitem{Simon}  B. Simon, {\sl Functional Integration and Quantum Physics}
(Academic Press, New York, 1979).

\bibitem{FeynStatMech}  R. P. Feynman, {\sl Statistical Mechanics: A Set of
Lectures} (W. A. Benjamin, Reading, MA, 1972).

\bibitem{TVS3}  F. Brosens, J. T. Devreese, and L. F. Lemmens, Phys. Rev. E 
{\bf 55}, 227 (1997).

\bibitem{Scatlens}  For the scattering lengths of the alkali gasses we used
the following sources: E. R. I. Abraham {\it et al.}, Phys. Rev. Lett. {\bf %
74}, 1315 (1995) for lithium; K. B. Davis {\it et al.}, Phys. Rev. Lett {\bf %
74}, 5202 (1995) and A. J. Moerdijk {\it et al.}, Phys. Rev. Lett. {\bf 73},
519 (1994) for sodium; N. R. Newbury {\it et al.}, Phys. Rev. A {\bf 51},
2680 (1995) for rubidium-87; Gardner {\it et al.}, Phys. Rev. Lett. {\bf 74}%
, 3764 (1995) for rubidium-85; C. R. Monroe {\it et al.}, Phys.\ Rev. Lett. 
{\bf 70}, 414 (1993) and M. Arndt {\it et al.}, Phys. Rev. Lett. {\bf 79},
625 (1997) for cesium.

\bibitem{Confine}  Typical confinement frequencies were taken from C. C.
Bradley {\it et al.}, Phys. Rev. Lett. {\bf 78}, 985\ (1997) for lithium; K.
B. Davis {\it et al.}, Phys. Rev. Lett {\bf 74}, 5202 (1995) for sodium; D.
S. Jin {\it et al.}, Phys. Rev. Lett. {\bf 78}, 764 (1997) for rubidium-87;
Gardner {\it et al.}, Phys. Rev. Lett. {\bf 74}, 3764 (1995) for
rubidium-85; M. Arndt {\it et al.}, Phys. Rev. Lett. {\bf 79}, 625 (1997)
for cesium.

\bibitem{Salas97}  L. Salasnic, Modern Physics Letters B {\bf 11}, 1249
(1997).

\bibitem{BraPRL78}  C. C. Bradley, C. A. Sackett, and R. G. Hulet, Phys.
Rev. Lett. {\bf 78}, 985\ (1997).

\bibitem{cotPRA94}  R. C\^{o}t\'{e}, A. Dalgarno, and M. J. Jamieson, Phys.
Rev. A {\bf 50}, 399 (1994).

\bibitem{KraPRL96}  W. Krauth, Phys. Rev. Lett. {\bf 77}, 3695 (1996).

\bibitem{Gruunp97}  P. Gruter, D. Ceperley, and F. Lalo\"{e}, Phys. Rev.
Lett. {\bf 79, }3549 (1997).

\bibitem{WilPRL98}  J. Williams, R. Walser, C. Wieman, J. Cooper, and M.
Holland, Phys. Rev. Letters, April 98.

\bibitem{HulAPB65}  C. A. Sackett, C. C. Bradley, M. Welling, and R. G.
Hulet, Appl. Phys. B {\bf 65}, 433 (1997).

\bibitem{BuPRL97}  E. A. Burt, R. W. Ghrist, C. J. Myatt, M.J. Holland, E.
A. Cornell, and C. E. Wieman, Phys. Rev. Lett. {\bf 79}, 337 (1997).

\bibitem{KagJETP85}  Y. Kagan, B.V. Shustunov, and G. V. Shlyapnikov, JETP
Lett. {\bf 42}, 209 (1985).

\bibitem{Kubo}  R. Kubo, {\sl Statistical Mechanics} (North-Holland,
Amsterdam, 1974).

\bibitem{TVS6}  L. F. Lemmens, F. Brosens, and J. T. Devreese, Solid State
Commun. {\bf 109}, 615 (1999).
\end{references}
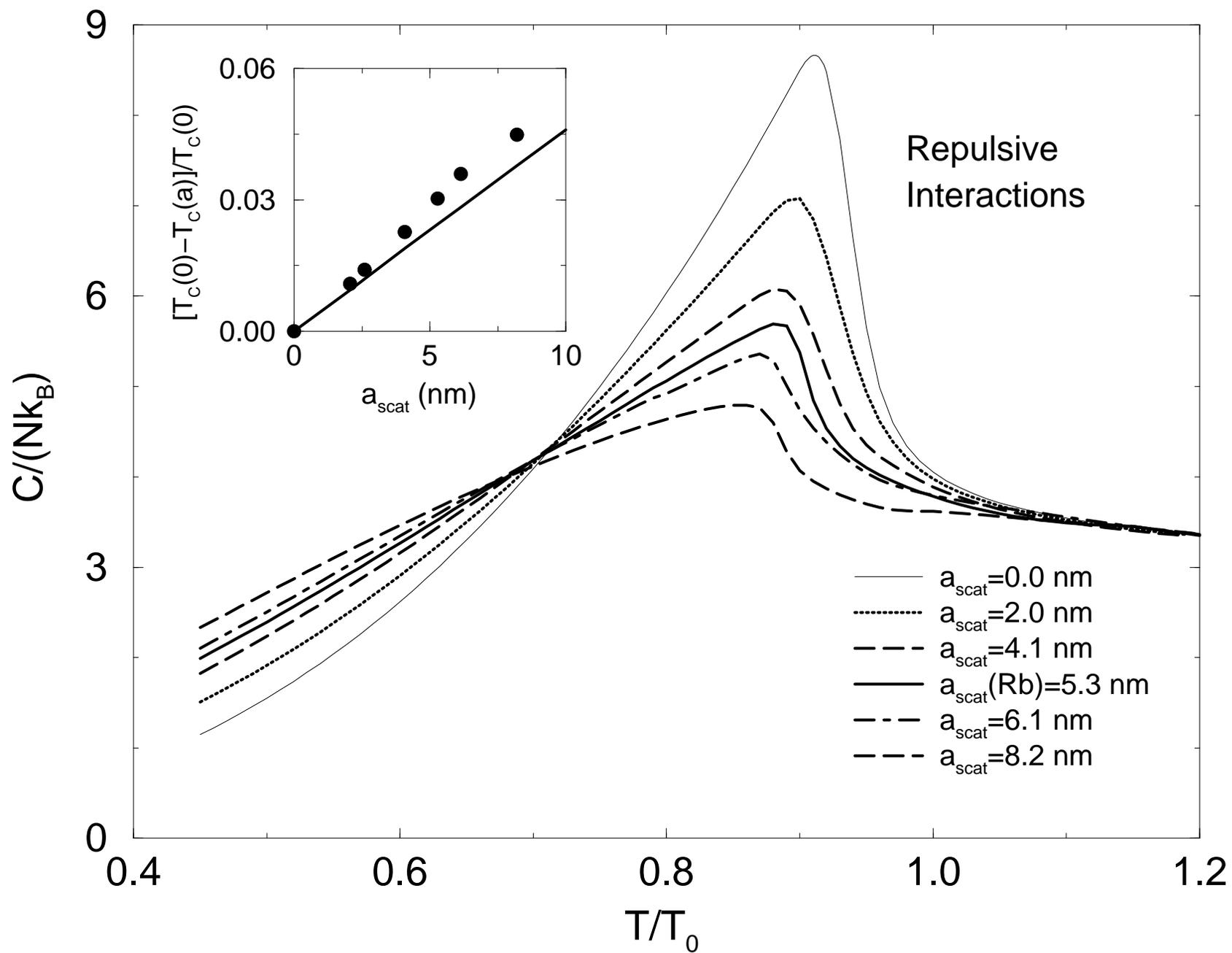
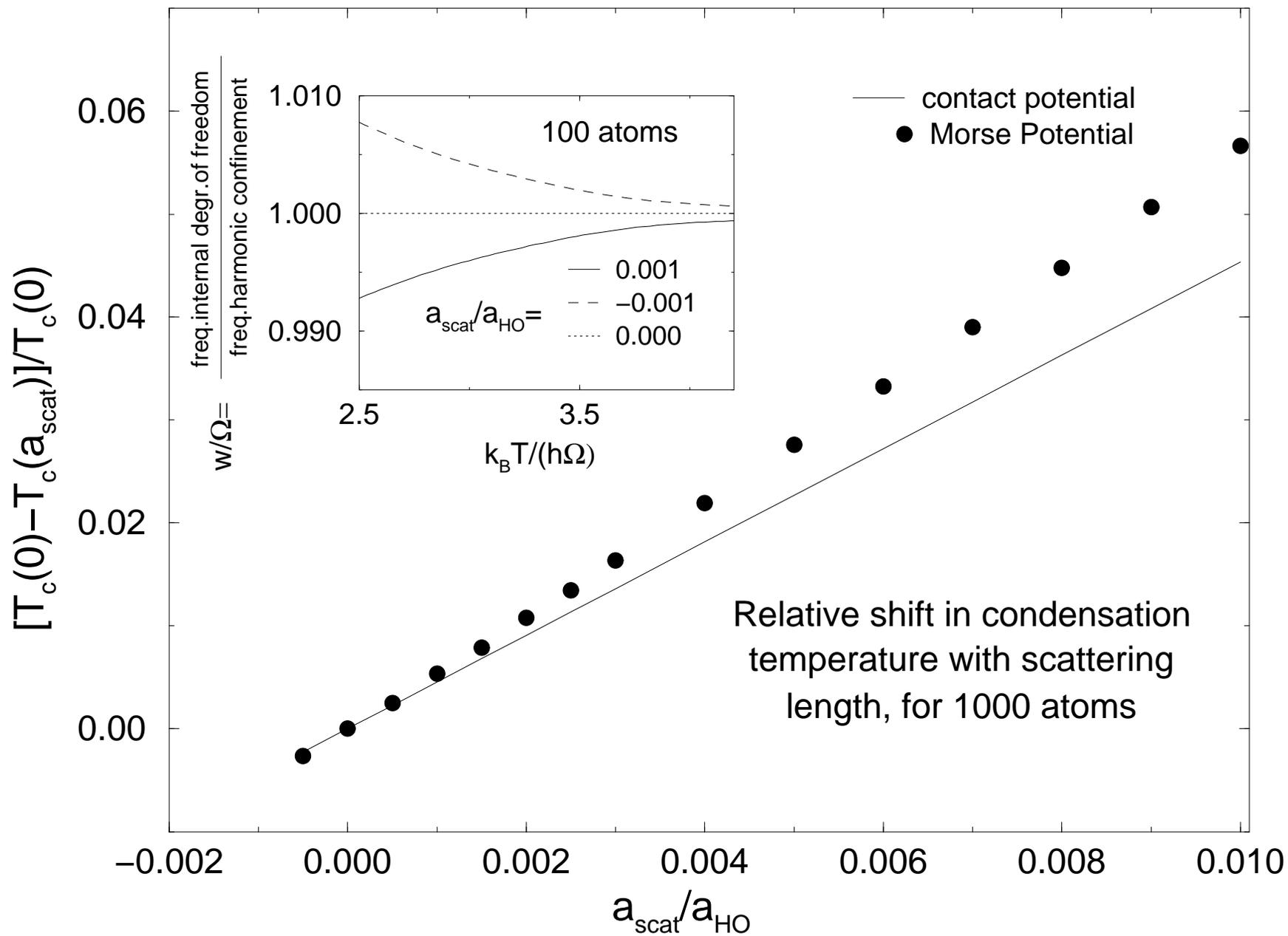
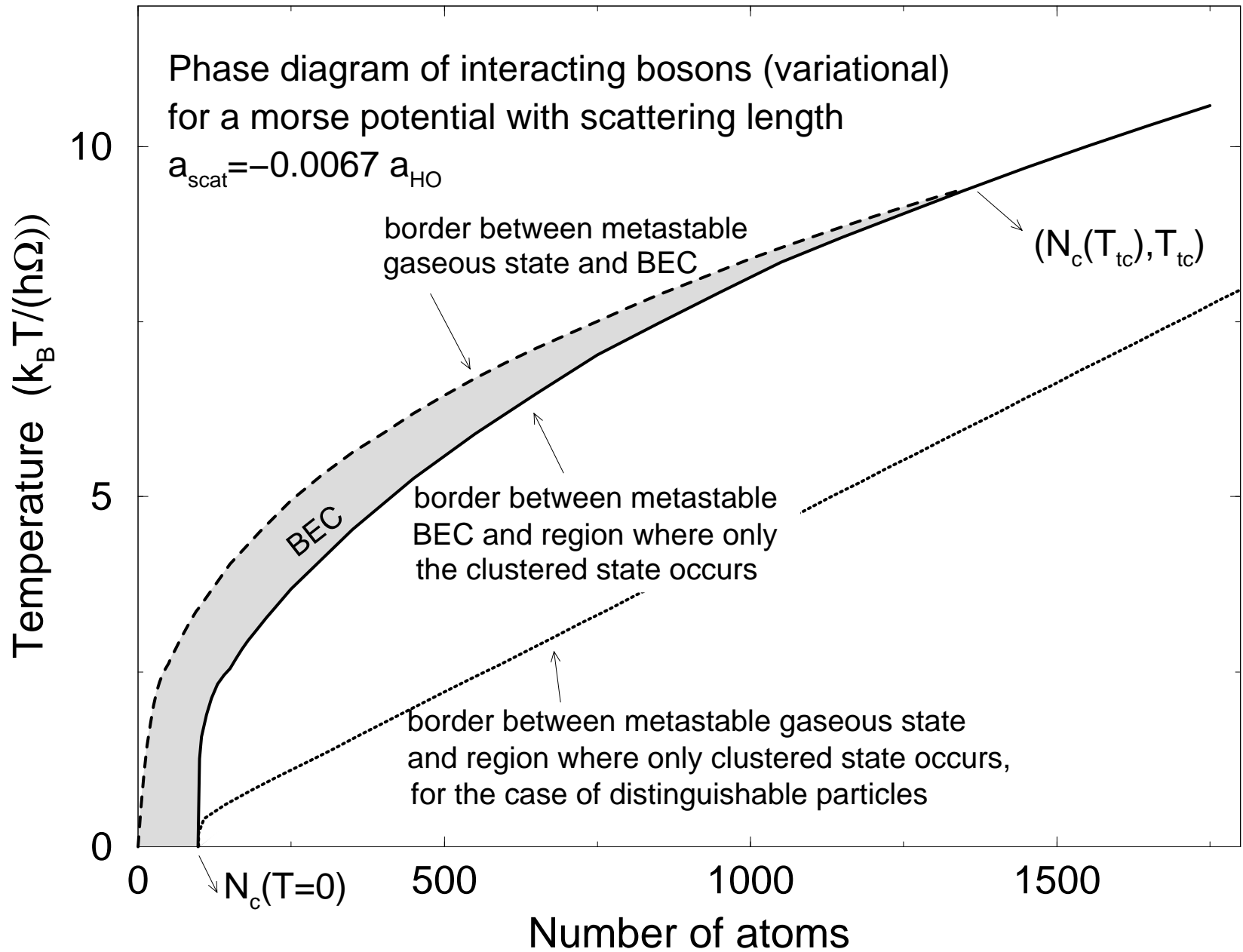
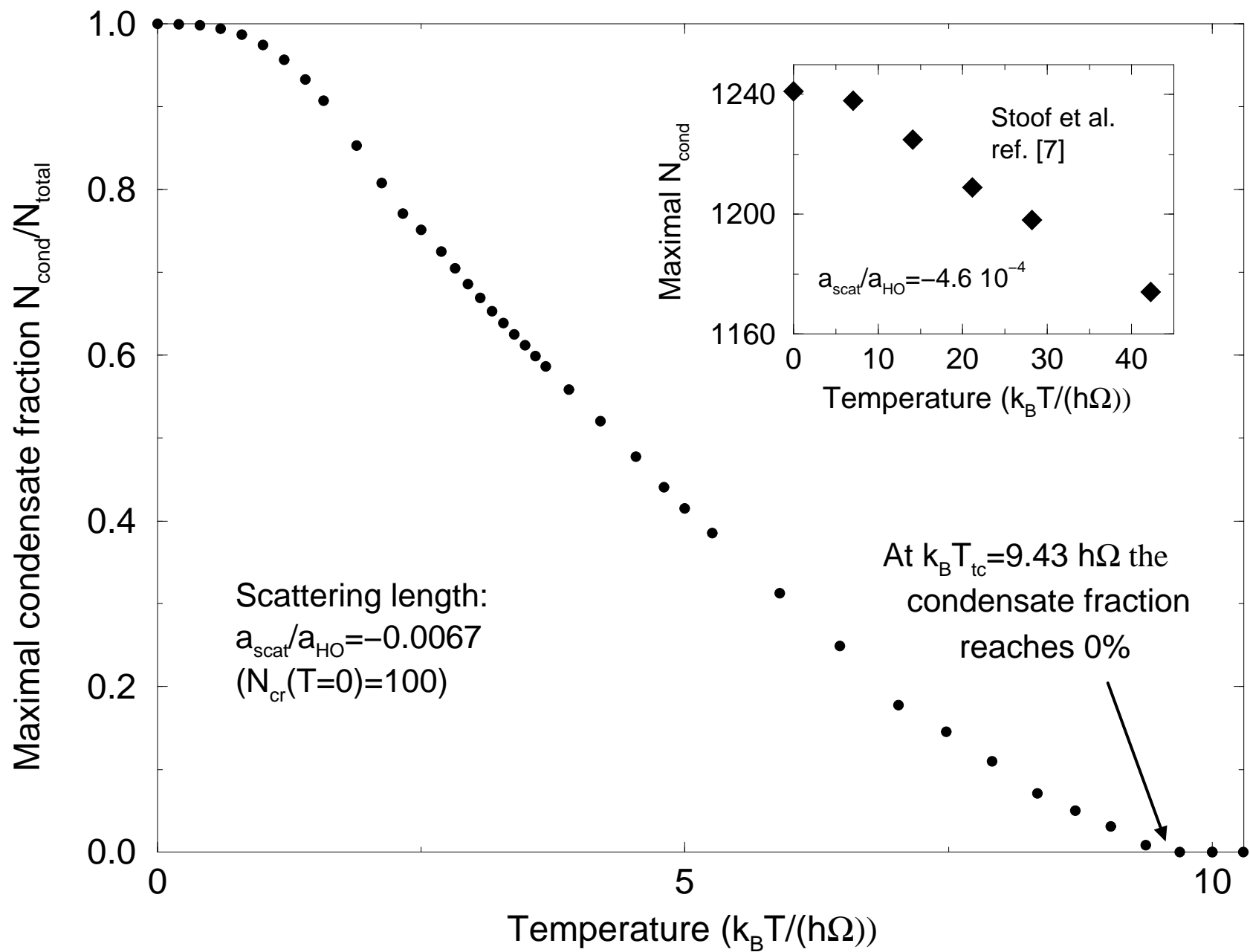
\end{document}